\documentclass[useAMS,usenatbib]{mn2e}

\usepackage{epsfig}
\usepackage{amsmath}
\usepackage{amssymb}
\usepackage{natbib}
\usepackage{threeparttable} 
\usepackage{booktabs}
\usepackage{multirow}
\usepackage{color}
\usepackage{hyperref}

\usepackage{epstopdf}
\usepackage{times}

\newcommand{\chan}{\textit{Chandra}}
\newcommand{\swift}{\textit{Swift}}

\newcommand{\xmm}{\textit{XMM-Newton}}

\newcommand{\maxi}{\textit{MAXI}}

\newcommand{\suzaku}{\textit{Suzaku}}
\newcommand{\nustar}{\textit{NuSTAR}}

\newcommand{\nicer}{\textit{NICER}}

\newcommand{\Msun}{\mathrm{M}_{\odot}}
\newcommand{\lum}{\mathrm{erg~s}^{-1}}
\newcommand{\flux}{\mathrm{erg~cm}^{-2}~\mathrm{s}^{-1}}

\newcommand{\nh}{\mathrm{cm}^{-2}}

\newcommand{\ledd}{$L_{\mathrm{Edd}}$}
\newcommand{\lx}{$L_{\mathrm{X}}$}
\newcommand{\rg}{$r_{\mathrm{g}}$}

\newcommand{\source}{4U~1608$-$52}
\def \mnras {MNRAS}
\def \apj {ApJ}
\def \apjs {ApJS}
\def \apjl {ApJL}
\def \aap {A\&A}
\def \nat {Nature}

\def \pasj {PASJ}

\title[Disappearing reflection in 4U 1608$-$52]{A strongly changing accretion morphology during the outburst decay of the neutron star X-ray binary \source}
\author[Van den Eijnden, Degenaar, Ludlam, et al.]
{J. van den Eijnden$^{1}$\thanks{e-mail: A.J.vandenEijnden@uva.nl}, N. Degenaar$^{1}$, R. M. Ludlam$^{2,\dagger}$, A. S. Parikh$^{1}$, J.M.~Miller$^{3}$, 
\newauthor R. Wijnands$^{1}$, K. C. Gendreau$^{4}$, Z. Arzoumanian$^{4}$, D. Chakrabarty$^{5}$, P. Bult$^{4,6}$ \\
$^1$ Anton Pannekoek Institute for Astronomy, University of Amsterdam, Science Park 904, 1098 XH, Amsterdam, the Netherlands\\
$^2$Cahill Center for Astronomy and Astrophysics, California Institute of Technology, Pasadena, CA 91125, USA, 
$^{\dagger}$ NASA Einstein Fellow\\
$^3$Department of Astronomy, University of Michigan, 1085 South University Avenue, Ann Arbor, MI  48109, USA\\
$^{4}$Astrophysics Science Division, NASA Goddard Space Flight Center, Greenbelt, MD 20771, USA\\
$^{5}$MIT Kavli Institute for Astrophysics and Space Research, Massachusetts Institute of Technology, Cambridge, MA 02139, USA\\
$^{6}$Department of Astronomy, University of Maryland, College Park, MD 20742, USA\\}

\begin{document}

\date{Accepted XXX. Received YYY; in original form ZZZ}

\pagerange{\pageref{firstpage}--\pageref{lastpage}} \pubyear{0000}

\maketitle

\label{firstpage}

\begin{abstract}
It is commonly assumed that the properties and geometry of the accretion flow in transient low-mass X-ray binaries (LMXBs) significantly change when the X-ray luminosity decays below $\sim 10^{-2}$ of the Eddington limit (\ledd). However, there are few observational cases where the evolution of the accretion flow is tracked in a single X-ray binary over a wide dynamic range. In this work, we use \nustar\ and \nicer\ observations obtained during the 2018 accretion outburst of the neutron star LMXB \source, to study changes in the reflection spectrum. We find that the broad Fe-K$\alpha$ line and Compton hump, clearly seen during the peak of the outburst when the X-ray luminosity is $\sim 10^{37}~\lum$ ($\sim 0.05$~\ledd), disappear during the decay of the outburst when the source luminosity drops to $\sim 4.5 \times 10^{35}~\lum$ ($\sim 0.002$~\ledd). We show that this non-detection of the reflection features cannot be explained by the lower signal-to-noise at lower flux, but is instead caused by physical changes in the accretion flow. Simulating synthetic \nustar\ observations on a grid of inner disk radius, disk ionisation, and reflection fraction, we find that the disappearance of the reflection features can be explained by either increased disk ionisation ($\log \xi \gtrsim 4.1$) or a much decreased reflection fraction. A changing disk truncation alone, however, cannot account for the lack of reprocessed Fe-K$\alpha$ emission. The required increase in ionisation parameter could occur if the inner accretion flow evaporates from a thin disk into a geometrically thicker flow, such as the commonly assumed formation of an radiatively inefficient accretion flow at lower mass accretion rates.
\end{abstract}


%%%%%%%%%%%%%%%%%
% INTRODUCTION
%%%%%%%%%%%%%%%%%

\section{Introduction}
\label{sec:introduction}
Low-mass X-ray binaries (LMXBs), in which a neutron star or a black hole attracts gas from a low-mass companion star, are prime tools to study the physics of accretion. The X-ray luminosity (\lx) of LMXBs scales with the rate at which matter is accreted by the compact object. The population of transient sources alternates outbursts of accretion with quiescent episodes and can therefore be studied over a wide range of X-ray luminosities; from the Eddington limit, \ledd $\simeq 10^{38} (M/\Msun)~\lum$, down to a very small fraction of it (\lx$\sim$$10^{-8}$~\ledd). 

%Although LMXBs have been known and extensively studied for over 5 decades, 
It is observationally established that at high X-ray luminosities ($\gtrsim$$10^{-2}$~\ledd), the accreted gas spirals in a thin Keplerian disk that typically extends close to the compact object \citep[e.g.,][for a review]{done2007}. At very low accretion rates ($\lesssim$$10^{-4}-10^{-5}$~\ledd), on the other hand, the accretion disk is likely truncated further away from the black hole or the neutron star \citep[e.g.,,][]{zdziarski1999,esin2001,kalemci2004,tomsick2004}. Nevertheless, it is currently unclear at what accretion luminosity the disk begins to move away, how this truncation proceeds with changing luminosity, and if this process differs for neutron stars and black holes. Standard accretion theory suggests that with decreasing accretion rates, an increasing part of the inner disk evaporates into a radiatively inefficient hot flow \citep[e.g.,][]{narayan1994, esin1997}. This classical disk truncation mechanism is expected to be more efficient for black holes, because for neutron stars the soft photons from the stellar surface can cool the hot flow and make it condense back on the disk \citep[][]{narayan1995}.

In neutron star LMXBs, a second mechanism that can lead to disk truncation may be at play; the stellar magnetic field may push the inner disk out \citep[e.g.,][]{ibragimov2009,miller2011,degenaar2014_groj1744,degenaar2017_igrj1706,vandeneijnden2017}. One may expect this effect to become larger with decreasing accretion rate, and hence lower pressure exerted by the disk. However, there is little observational data to test this idea in a single source \citep[][]{vandeneijnden2018_igr}. The interaction between the stellar magnetic field and accretion flow plays an important role in a number of key processes, including the production of outflows \citep[e.g.,][]{romanova2009,deller2015,vandeneijnden2018_igr}, and the spin evolution of neutron stars \citep[e.g.,][]{ghosh1978,alpar1982,wijnands1998_timing1808,dangelo2017,bhattacharyya2017_spin}. Yet, while the interaction between the accretion disk and the magnetosphere is clearly important for the evolution and appearance of neutron stars, it remains poorly understood.

X-ray reflection modelling is a powerful tool to constrain the geometry of accretion flows in LMXBs. This X-ray emission -- originating from close to the accretor and then reprocessed or `reflected' toward the observer by the accretion disk -- manifests itself most prominently as an Fe-K$\alpha$ emission line at $\sim$6--7~keV and the Compton hump at $\sim$20--40 keV. The shape of these features is modified by Doppler and gravitational redshift effects as the gas in the disk moves in high-velocity Keplerian orbits inside the gravitational well of the compact accretor. In particular, reflection modelling allows for a measure of the inner radial extent of the accretion disk, $R_{\mathrm{in}}$. The reflection spectrum thus encodes information about the morphology of the accretion flow \citep[e.g.,][for reviews]{fabian2010,dauser2016}. In addition, the reflected spectrum is affected by the internal properties of the disk, such as level of ionisation and composition. 

Comparing the reflection spectrum at different fractions of the Eddington accretion rate in the same source can reveal changes in the accretion flow structure and properties as the X-ray luminosity decays. An example of a source where the reflection spectrum has been tracked over a range in X-ray luminosity is the black hole LMXB GX 339$-$4 \citep[e.g.,][]{garcia2015_gx339m4}. In this source, the relation between X-ray luminosity and disk inner radius has been determined in detail, showing a general trend of decreasing inner radius with Eddington fraction \citep{miller2006_gx339m4,reis2008,tomsick2009,petrucci2014,garcia2015_gx339m4,wang-ji2018,garcia2019}. This large set of measurements is possible due to high duty cycle of GX 339$-$4, yielding over twenty measurements of the reflection spectrum in the past fifteen years. 

However, such systematic studies have to date been rare, especially in neutron stars, given the deep exposures required to detect reflection or otherwise obtain constraining upper limits at low luminosities (e.g., \lx$< 10^{-2}$~\ledd). In addition, the latter luminosity regime is often traversed quickly and unpredictably, complicating the scheduling of observational campaigns in sources that show fewer outbursts. Finally, for neutron stars, a mere measurement of the inner radius does not necessarily reveal the driving factor of the disk truncation: the formation of a radiatively inefficient hot flow, truncation by the magnetic field, or possibly both \citep{cackett2008_iron,degenaar2017_igrj1706}. So while \textit{NuSTAR} reflection spectra have been measured for many individual neutron star LMXBs \citep[see][for a recent overview]{ludlam2019}, differences between their neutron star properties mean variations in the reflection spectrum can not be directly attributed to accretion rate. For such an inference, a reflection study of a single source at different accretion rates is essential.  

Given these challenges for neutron star LMXBs, \source\ \citep[][]{grindlay1976,tananbaum1976} is a promising source to track the reflection spectrum and accretion geometry at different outburst stages, down to \lx$< 10^{-2}$~\ledd. Firstly, a \nustar\ observation performed during its 2014 outburst, when it was accreting at $\sim10^{-2}$~\ledd\ ($\sim 10^{36}~\lum$), revealed very strong reflection features that allowed to measure an inner disk radius of $8.5 \pm 1.5$~\rg \citep[\rg$=GM/c^2$][]{degenaar2015_4u1608}. The strength of its reflection is a very important advantage over other frequently active neutron star LMXBs, such as Aql X-1 and SAX J1808.4--3658, for which the reflection spectrum is much weaker \citep[e.g.,][]{cackett2008_iron,ludlam2017_aqlx1}. 
Secondly, \source\ is relatively close \citep[$\sim 3-4.5$~kpc;][]{galloway2008,poutanen2014_4u1608}, so that a large flux is obtained even when it accretes at a low Eddington luminosity. Thirdly, \source\ is among the most active transient neutron star LMXBs, going into outburst about once every 2--4 yr \citep[e.g.,][]{lochner1994,simon2004}, easing the scheduling of observations. Finally, the outburst profiles are often similar, making it feasible to plan observations at a specific X-ray luminosity.

In this work, we report on our efforts to use reflection spectroscopy to constrain the changing accreting morphology in the neutron star LMXB \source~in its 2018 outburst, using the \textit{NuSTAR}, \textit{NICER}, and \textit{Swift} X-ray observatories.

%%%%%%%%%%%%%%%%%
% OBSERVATIONS
%%%%%%%%%%%%%%%%%

\section{Observations and data analysis}

\subsection{Set up of the observation program}

The \maxi\ transient alert system \citep[][]{negoro2016} signalled activity from \source\ on 19 June 2018 June 19 (MJD 58288; nova ID 8288849999). When \maxi\ monitoring indicated that \source\ had evolved beyond its outburst peak, we started to monitor the source with the \textit{X-ray telescope} (XRT) aboard the \textit{Neil Gehrels Swift} observatory \citep[\swift;][]{gehrels2004}. This \swift\ monitoring was set up to facilitate triggering a \nustar\ observation (PI Degenaar), aiming to catch the source at a luminosity of $\sim 10^{35}~\lum$ ($\sim10^{-3}$~\ledd), i.e. a factor $\sim$10 lower than the 2014 observation of the source. In addition to our \swift\ and \nustar\ program, we used observations obtained with \nicer\ \citep[][]{gendreau2016} to investigate if, and how, the profile of the Fe-K$\alpha$ line changed along the course of the 2018 outburst. %The \maxi\ monitoring light curve \citep[2--20 keV;][]{maxi2009} is shown in Figure~\ref{fig:swift}.

\subsection{Spectral modelling}
All spectral fits reported in this work were performed using \textsc{XSpec} \citep[v. 12.10.1;][]{xspec}. To account for interstellar absorption along our line of sight, we used the model \textsc{tbabs}, with \textsc{vern} cross-sections \citep[][]{verner1996} and \textsc{wilm} abundances \citep[][]{wilms2000}, in our spectral modelling. Reported errors for spectral parameters indicate 1$\sigma$ confidence levels. To calculate luminosities from measured fluxes, we adopted a distance of 3.6 kpc.

\subsection{\swift}

To put the \nustar\ and \nicer\ observations in perspective, we use 22 \swift/XRT observations (obsID 00010741001--26) to illustrate the source's evolution along the 2018 outburst. These observations were obtained between 3 July and 24 October 2018 (MJD 58302 and 58415) and had typical exposure times of $\sim$1~ks. We followed standard data reduction procedures, using \textsc{xselect} to extract spectra and \textsc{xrtmkarf} to create arf files, while the rmf file was taken from the \textsc{caldb}. All XRT spectra were fitted to a simple absorbed powerlaw model (\textsc{tbabs*pegpwlw} in \textsc{XSpec}). In these fits, we fixed the hydrogen column density at $N_{\mathrm{H}} = 1.58 \times 10^{22}~\nh$ (as determined from a simultaneous fit to all XRT spectra where $N_{\mathrm{H}}$ was tied). The unabsorbed 0.5--10 keV flux was subsequently determined using the convolution model \textsc{cflux}. While in the main paper we only show the XRT flux light curve (cf. Figure \ref{fig:swift}), details on these spectral fits can be found in Appendix \ref{app:XRT}. 

\subsection{\nustar}
\source\ was observed for $\sim$53~ks with \nustar\ on July 12--14 2018 (obsID 80401321002), providing data in the 3--79 keV energy band. We processed the data with \textsc{nupipeline}, incorporated in \textsc{heasoft} (v. 6.23), using \textsc{SAAMODE=optimized}. We extracted light curves and spectra for the two co-aligned grazing incidence hard X-ray imaging telescopes, with focal plane modules (FPM) A and B, using \textsc{nuproducts}. A circular region with a radius of $60''$ was used to extract source events and a void region of the same size, placed on the same detector, was used to obtain background events. 

All spectral data were grouped into bins with a minimum of 20 photons using \textsc{grppha}. \source\ was detected significantly above the background in the entire \nustar\ bandpass. Furthermore, when fitting the data of the two mirrors simultaneously with a freely-floating constant factor in between, we found the difference between the FPMA and B to be $<$1\% for this data set. Therefore, we summed the two spectra using \textsc{addascaspec} and in the following report only on the results obtained using this summed spectrum.  In this work, we compare this 2018 data set to the \nustar\ observation performed in 2014. For details on the 2014 observation and data reduction, we refer to \citet{degenaar2015_4u1608}.

\subsection{\nicer}

We used 12 \nicer\ observations (obsIDs 1050070125--1050070136) obtained between 21 June and 9 July 2018, and having typical exposure times of 0.1--2.5 ks. No Type-I X-ray bursts occurred during the \nicer\ monitoring. The data were reduced using {\sc nicerdas} 2018-10-07\_V005 with the standard \nicer\ filtering. Additionally, we created GTIs to select a SUN\_ANGLE $>60^{\circ}$, COR\_SAX $>4$, and KP $<5$ via the {\sc nimaketime} tool. The latter two corrections define the cut-off magnetic rigidity and the space weather conditions. We applied the GTIs to the data using the {\sc niextract-events} command. A light curve was then extracted by loading the cleaned event files into {\sc xselect}. We separate the \nicer\ monitoring of the 2018 outburst into five intervals shown in Figure \ref{fig:nicer}. A spectrum was were extracted from each interval.

Because \nicer\ does not have imaging capabilities, backgrounds were generated from periodic observations of RXTE blank sky field locations \citep{jahoda2006}. We extracted a time-averaged background from 51 observations of blank sky field 5 that were reduced in the same manner as \source. The background spectrum was subtracted from the source spectra prior to normalizing the data to the Crab Nebula \citep[see][for more details]{ludlam2018}. This accounts for instrumental uncertainties within the data as \nicer's calibration continues to be refined. We note that the source was significantly detected above the background in all used observations.

 \begin{figure}
 \begin{center}
	\includegraphics[width=8.5cm]{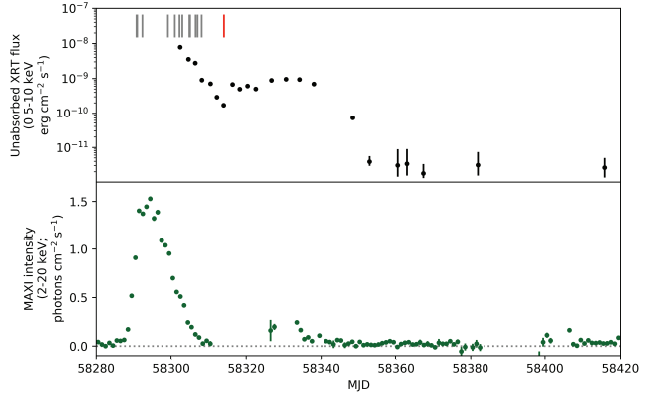}
    \end{center}
    \caption[]{\swift/XRT (top, 0.5--10 keV, binned per observation) and \maxi\ (bottom, 2--20 keV, binned per day) light curves of the 2018 outburst of \source. The grey and red bars in the top panel indicate the times of our \nicer\ and \nustar\ observations, respectively.
        }
 \label{fig:swift}
\end{figure}

 \begin{figure}
 \begin{center}
	\includegraphics[width=8.5cm]{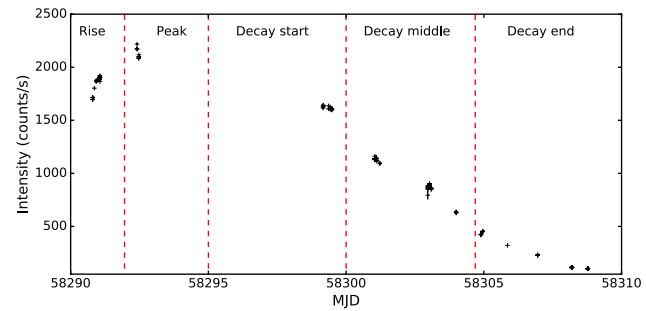}
    \end{center}
    \caption[]{\nicer\ count rate light curve of the 2018 outburst of \source\ (0.5--6.8 keV, binned by 128 s per observation). The dashed red lines indicates the different phases of the outburst for which we extracted X-ray spectra (presented in Figure~\ref{fig:ratio}).}
 \label{fig:nicer}
\end{figure}

 \begin{figure}
 \begin{center}
	\includegraphics[width=8.5cm]{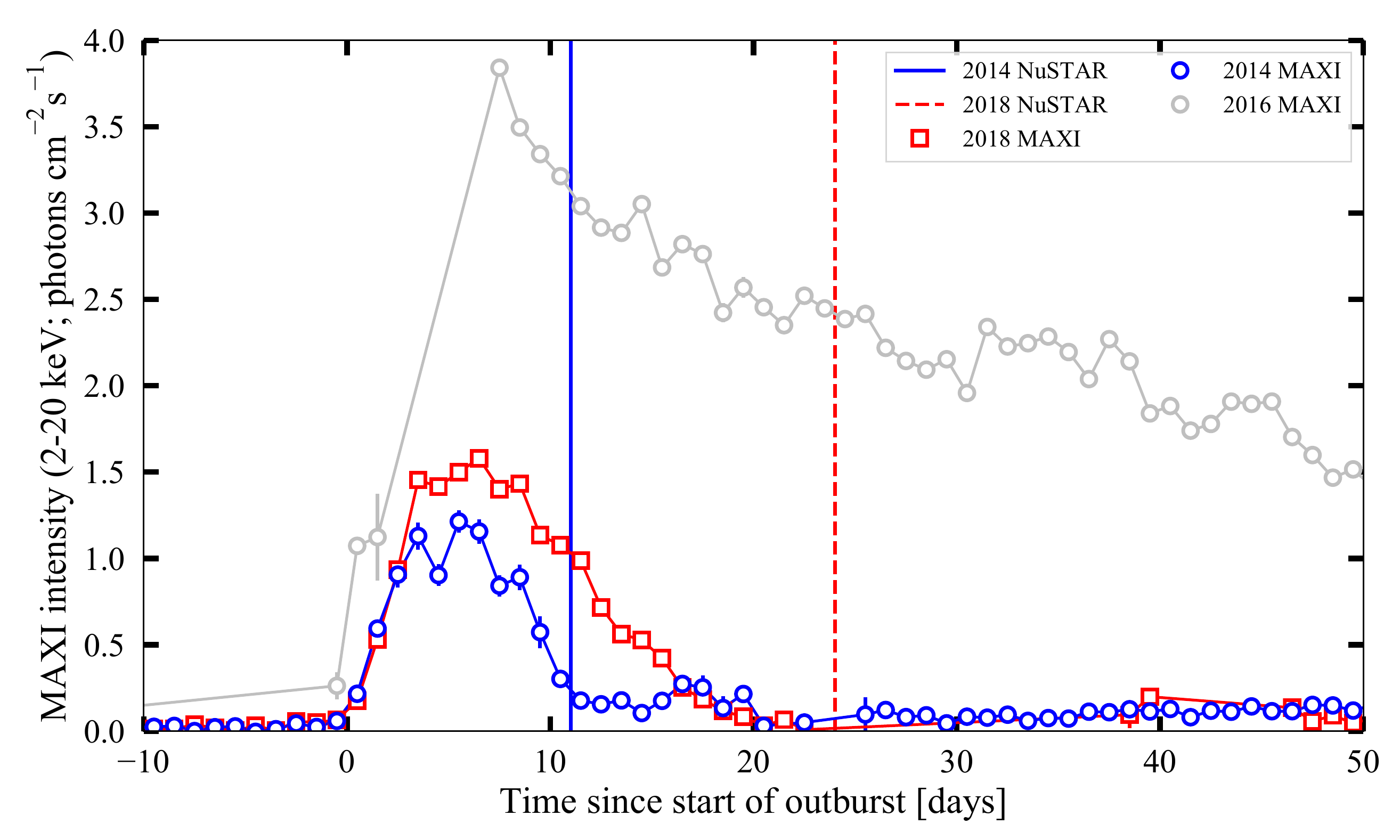}
    \end{center}
    \caption[]{\textit{MAXI} light curves of the 2014 and 2018 outbursts, shown in red and blue, respectively. The grey light curve shows the 2016 outburst, comfirming the similar faint/short nature of the 2014 and 2018 outbursts. The vertical lines indicate the relative times of the \textit{NuSTAR} observations in their respective outbursts. Note that the source flux during the 2018 \textit{NuSTAR} observation is below the \textit{MAXI} sensitivity.}
 \label{fig:maxi_comparison}
\end{figure}

%%%%%%%%%%%%%%%%%
% RESULTS
%%%%%%%%%%%%%%%%%

\section{Results}

The \swift/XRT light curve in Figure \ref{fig:swift} (top) shows how \source\ initially decayed steadily over a timescale of $\sim 10$~days, after which the 0.5--10 keV flux rose by a factor of a few, remaining steady for $\sim$20~days until finally returning to quiescence $\sim$2 months after the outburst peak. We note that the 0.5--10 keV flux detected in the last set of \swift\ observations is around the quiescent level reported from \chan\ observations \citep[][]{marino2018}. The 2018 \nustar\ observation was obtained exactly in the valley of the initial decay of the outburst (red bar in Figure~\ref{fig:swift} top). At this time, the 0.5--10 keV unabsorbed flux measured by \swift\ (Obsid 00010741006) was $(2.9 \pm 0.15) \times 10^{-10}~\flux$ (see Appendix \ref{app:XRT} for details on the \swift\ spectral fits). This translates into a luminosity of $(4.5 \pm 0.2) \times 10^{35}~\lum$ which is a factor $\sim$10 lower, as intended, than during the \nustar\ observation that was obtained during the 2014 outburst \citep{degenaar2015_4u1608}. 

\source\ shows regular outbursts, that broadly fall into two categories: relatively faint and short outbursts, and significantly brighter and longer ones. As the two \textit{NuSTAR} observations were taken in two outbursts, we compare their profiles in Figure \ref{fig:maxi_comparison}. While slight differences exist in peak count rate and duration, the comparison with a 2016 outburst (shown in grey) shows that both outbursts clearly fall in the faint/short category. Therefore, in the remainder of this paper, we will assume the accretion geometry follows a similar evolution throughout both outburst decays. The comparison of the 2018 \textit{NICER} and the 2014 \textit{NuSTAR} observations support this assumption, as shown in the next Section. 

\subsection{The disappearance of reflection during the decay}\label{subsec:disappear}

\begin{figure*}
 \begin{center}
	\includegraphics[width=8.5cm]{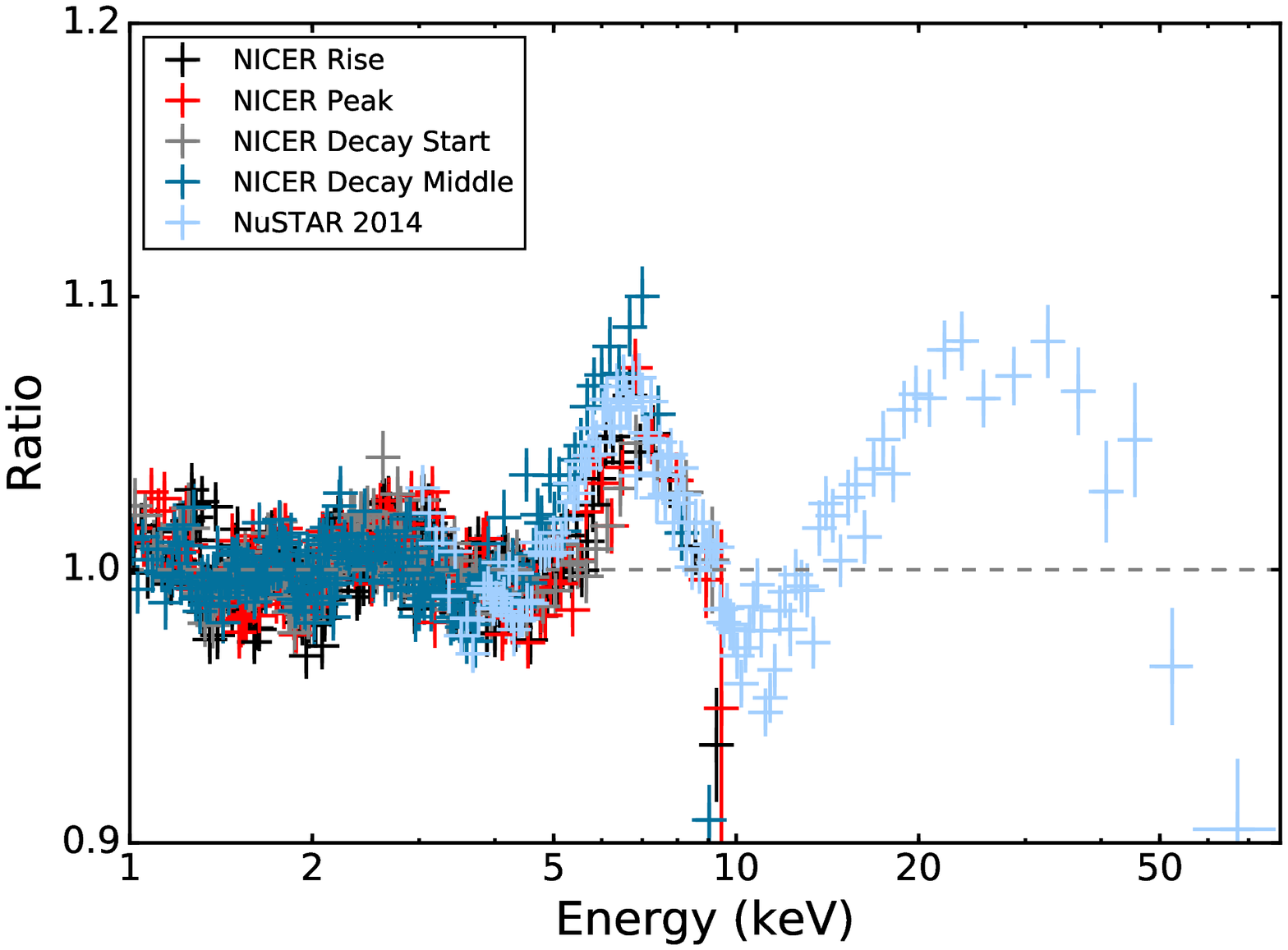}\hspace{+0.0cm}
		\includegraphics[width=8.5cm]{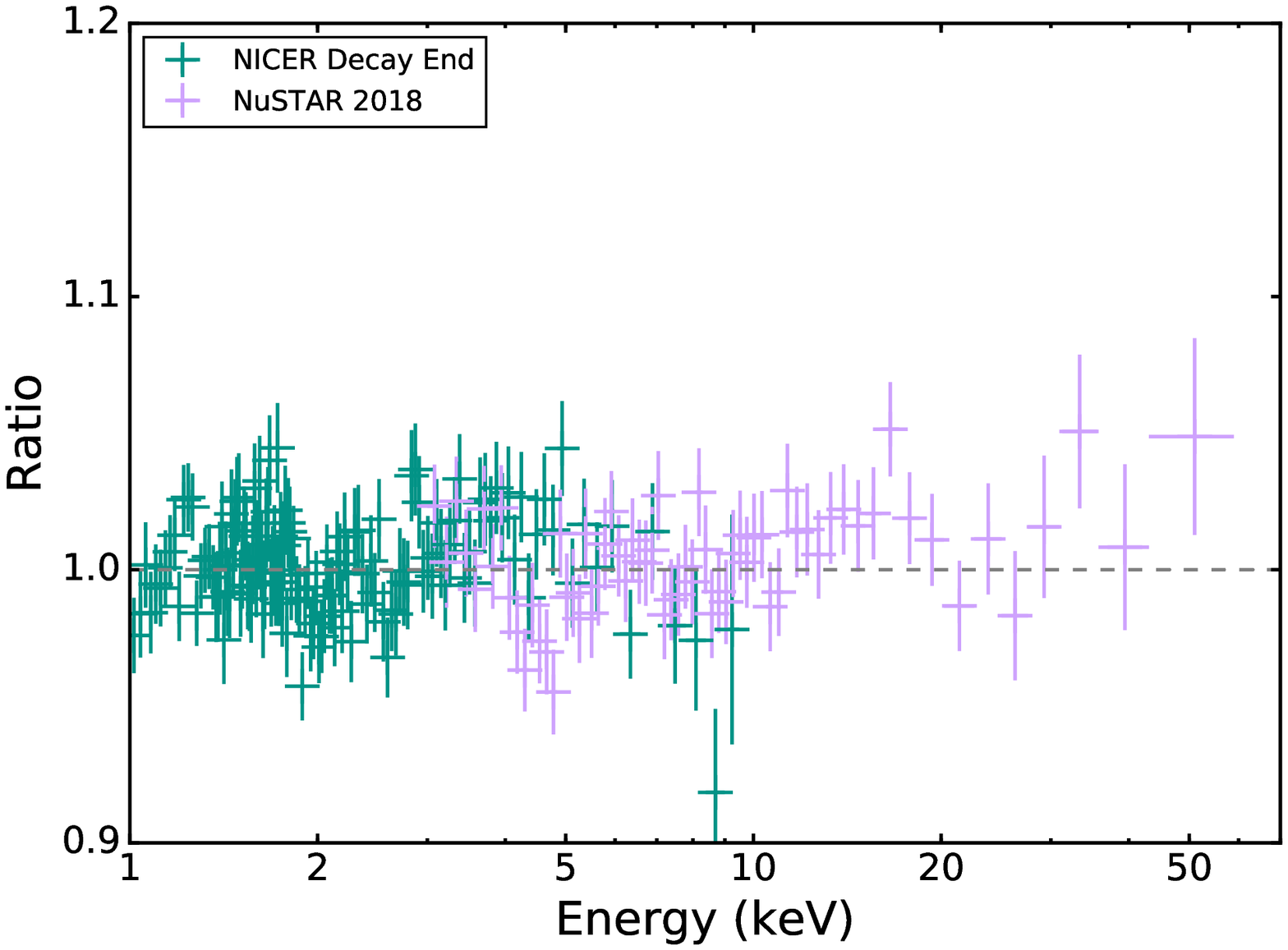}
    \end{center}
    \caption[]{Data to model ratio after fitting the data to a simple phenomenological model (see text for details). Left: \nicer\ data obtained during different epochs of the 2018 outburst (see Figure~\ref{fig:nicer}), compared to \nustar\ data obtained during the 2014 outburst. This plot illustrates that an Fe-K$\alpha$ line profile is clearly seen during the main part of the 2018 outburst, similar to that seen in 2014. Right: Comparison of  \nicer\ data obtained at the end of the decay of the 2018 outburst (see Figure~\ref{fig:nicer}) with the  \nustar\ observation taken around this time. This plot illustrates i) the sudden disappearance of the Fe-K$\alpha$ line profile in the \nicer\ data, and ii) the similarity between the spectral shape in the \nicer\ and \nustar\ data.}
 \label{fig:ratio}
\end{figure*}

During the 2014 outburst peak, \nustar\ revealed strong reflection features in \source, detecting both a broad Fe-K$\alpha$ line and a Compton hump \citep{degenaar2015_4u1608}. With \textit{NICER} monitoring, we can assess whether such features were also initially present during the 2018 outburst. For this purpose, we initially performed a quick-look analysis of both the \nicer\ spectra of the first four epochs (rise, peak, decay start, middle; see Figure \ref{fig:nicer}) and the archival \textit{NuSTAR} spectrum using a simple continuum model, \textsc{tbabs}$\times$(\textsc{bbodyrad}$+$\textsc{nthcomp}). To highlight the reflection features and only fit the continuum, the $5$--$7$ keV and $15$--$30$ keV ranges were ignored. In the left panel of Figure~\ref{fig:ratio}, we show the data-to-model ratio for these spectral fits in the $1$--$79$ keV energy range. We can see clearly that during the rise, peak, and early/middle decay of the outburst, a strong, broad Fe-K$\alpha$ line is present with a stable shape similar to the line profile seen in the 2014 \nustar\ observation. 

Repeating the same quick-look exercise with the spectrum of the final \textit{NICER} epoch, and the 2018 \textit{NuSTAR} observation (and using a more simple \textsc{tbabs}$\times$\textsc{nthcomp} model due to the lower flux), we can see that both the Fe-K$\alpha$ line and the Compton hump disappear. We show this evolution of the reflection spectrum in the right panel of Figure \ref{fig:ratio}.

The disappearance of the Fe-K$\alpha$ line, combined with its remarkable prior stability, as observed by \textit{NICER} is suggestive of a distinct change in the accretion flow geometry and/or intrinsic properties on a days time scale. We choose, however, not to fit physical reflection models to the \nicer\ data, for two reasons: firstly, \nicer\ spectral calibration continues, at the time of writing, to be refined. Secondly, and importantly, the soft \nicer\ bandpass does not include the Compton hump clearly seen in the 2014 \nustar\ spectrum (Figure \ref{fig:ratio}, left panel). As this difference in coverage limits the ability to constrain the parameters of the reflection spectrum, we instead focus on the deep 2018 \nustar\ observation in the remainder of this work. 

\begin{figure}
	\includegraphics[width=\columnwidth]{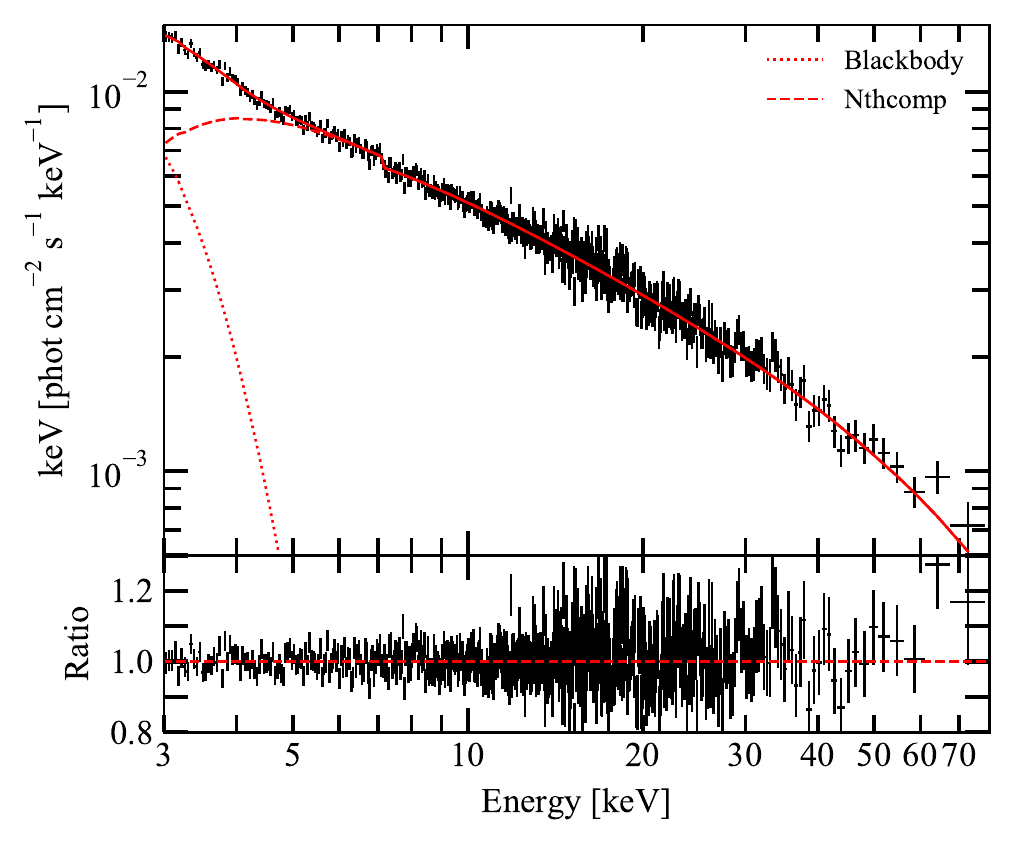}
    \caption[]{\textit{Top:} The 2018 \textit{NuSTAR} spectrum of 4U 1608$-$52, unfolded around the best-fitting \textsc{tbabs}$\times$(\textsc{bbodyrad}$+$\textsc{nthcomp}) continuum model using the \textsc{xspec} command \textsc{eufspec}. The data have been visually rebinned for clarity to a S/N$>10$. The red dotted, dashed, and solid lines show the thermal, Comptonized, and total model, respectively. \textit{Bottom:} data-to-model ratio for the continuum model shown in the top panel.}
 \label{fig:2018spectrum}
\end{figure}

For our detailed analysis, we start by fitting the 2018 \nustar\ spectrum with a simple phenomenological continuum model, \textsc{tbabs*(bbodyrad+nthcomp)}, where we tie the temperature of the \textsc{nthcomp} thermal input spectrum to that of the \textsc{bbodyrad} component. We fit the data in the full $3$--$79$ keV \nustar\ bandpass. This simple model, not containing any reflection features, provides an adequate description of the data, with a reduced $\chi^2$ of $0.89$ for $870$ degrees of freedom. In Figure \ref{fig:2018spectrum}, we show the spectrum and this continuum model, while the fitted spectral parameters are listed in Table \ref{tab:2018spectrum}. The data-to-model ratio shown in the bottom panel of Figure~\ref{fig:2018spectrum} confirms the lack of detected reflection features seen in the final \nicer\ epoch.  

\begin{table}
\caption{Fits to the 2018 \textit{NuSTAR} observation of 4U 1608$-$52. The continuum model consists of a \textsc{tbabs}$\times$(\textsc{bbodyrad}$+$
\textsc{nthcomp}) model, where the blackbody temperature $kT_{\rm BB}$ is linked between the thermal and comptonized component. In the reflection model, the \textsc{nthcomp} component is replaced by a \textsc{relxill} model, which contains both the Comptonized continuum and the broad-band reflection spectrum. In all reflection modelling, we fix the emissivity index, outer disk radius and cutoff energy to their default values, and set the spin parameter to $a=0.29$, corresponding to the known spin frequency. All uncertainties are given at the $1\sigma$ level.}
\begin{tabular}{lcc}
\label{tab:2018spectrum}
Parameters & \multicolumn{2}{c}{2018} \\
& Continuum & Reflection \\
\hline
\hline
$N_H$ [$10^{22}$ cm$^{-2}$] & $7.0\pm1.3$ & \\
$kT_{\rm BB}$ [keV] & $0.40 \pm 0.01$ & $0.43 \pm 0.02$  \\
$N_{\rm BB}$ & $(8.3^{+3.1}_{-2.3})\times10^2$ & $(4.4^{+4.9}_{-1.8})\times10^2$ \\
\hline
$\Gamma$ & $1.86 \pm 0.01$ & -- \\
$kT_e$ [keV] & $26.5^{+4.8}_{-3.2}$ & -- \\
$kT_{\rm BB}$ [keV] & $0.40 \pm 0.01$* & --  \\
$N_{\rm nthcomp}$ & $(1.7 \pm 0.1)\times10^{-2}$ & -- \\
\hline
Inclination [$^{\rm o}$] & -- & $56^{+14}_{-29}$ \\
$R_{\rm in}$ [\rg] & -- & $62^{+538}_{-56}$ \\
$\Gamma$ & -- & $1.75^{+0.02}_{-0.07}$ \\
$\log \xi$ & -- & $4.0^{+0.4}_{-0.2}$ \\
$A_{\rm Fe}$ & -- & $0.5^{+5.0}_{-0.5}$ \\
$F_{\rm ref}$ & -- & $1.4^{+1.6}_{-1.4}$ \\
$N_{\rm relxill}$ & -- & $(3 \pm 1)\times10^{-4}$ \\ \hline
$\chi^2_{\nu}$ (dof) & $0.895$ ($870$) & $0.903$ ($866$) \\ 
\hline
\end{tabular}
\end{table}

While no reflection features appear to be present, we also attempt a fit with a full reflection model, \textsc{tbabs*(bbodyrad+relxill)}. This model is based on that used by \citet{degenaar2015_4u1608} to fit the 2014 \nustar\ observation. The \textsc{relxill} model (version v1.0.4) includes both a power law continuum component and a fully relativistic, broad-band reflection model. The parameters that set the reflected spectrum are the emissivity profile, the neutron star spin, the system's inclination, inner and outer disk radius, ionisation parameter of the disk, iron abundance, cutoff energy, and reflection fraction (defined as the ratio of reflected and direct flux between 20 and 40 keV). The illuminating flux is characterised by the power law index. Following \citet{degenaar2015_4u1608}, we set the spin parameter to $0.29$, and freeze the outer disk radius to $400$ gravitational radii \rg. Additionally, we freeze the emissivity profile and the cutoff energy to their default values. Fitting this model does not yield a better fit than the continuum model, with a similar reduced $\chi^2$ of $0.90$ for $866$ degrees of freedom (i.e. four additional free parameters). Unsurprisingly, given the lack of reflection features, the \textsc{relxill} parameters are also poorly constrained. All parameters and fluxes with their uncertainties can be found in Table \ref{tab:2018spectrum}.

\subsection{Could reflection be hiding in the noise?}
\label{sec:results_isitburied}

No reflection features are detected during the last \nicer\ epoch and the 2018 \nustar\ observation. However, the lower flux raises the question whether the lack of detected reflection might arise from a lower signal-to-noise instead of physical changes in the source. To assess this question, we simulated synthetic \nustar\ observations: as a first step, we fit the \textsc{tbabs*(bbodyrad+relxill)} model to the 2014 \textit{NuSTAR} spectrum, reproducing the model fit that was reported in \citet{degenaar2015_4u1608}. We then scale the normalisations to the flux during the 2018 \nustar\ observations, and use the \textsc{xspec} \textsc{fakeit} command to simulate a spectrum with the same exposure time as the 2018 observation. We then fit this simulated spectrum with both the reflection and the continuum models, and compare those fits to test whether the 2014-level reflection would have been detected at the lower flux and 2018 exposure time.

The fit with the (correct) reflection model yields $\chi^2_{\nu} = 1343.6/1283 = 1.05$, while the continuum model returns a worse fit with $\chi^2_{\nu} = 1550.8/1287 = 1.20$. An f-test comparison of these models yields a chance probability of such an improvement by the reflection model of $p = 10^{-38}$. However, this does not take the change in continuum shape at lower flux into account. Therefore, we repeated this test by simulating a second fake spectrum from the reflection model, now using the continuum parameters (power law index, \textsc{relxill} normalisation, and blackbody temperature) as found for the 2018 \nustar\ data. For this second synthetic observation, the reflection and continuum models yielded $\chi^2_{\nu} = 1250.3/1283 = 0.97$ and $\chi^2_{\nu} = 1305.4/1287 = 1.01$ respectively. While the difference between the two fits is smaller, the (correct) reflection model still fits significantly better at an f-test probability of $2\times10^{-11}$. In other words, both simulations show that the 2018 \nustar\ observational setup should have detected reflection features as present during the 2014 (and 2018, as shown by \textit{NICER}) outburst peaks.  

Therefore, we conclude that the disappearance of the reflection features at lower flux is not only due to the decrease in signal-to-noise, but that physical changes in the accretion flow must also contribute. 

\subsection{Constraining the accretion geometry in the absence of reflection}

\begin{figure*}
 \begin{center}
	\includegraphics[width=\textwidth]{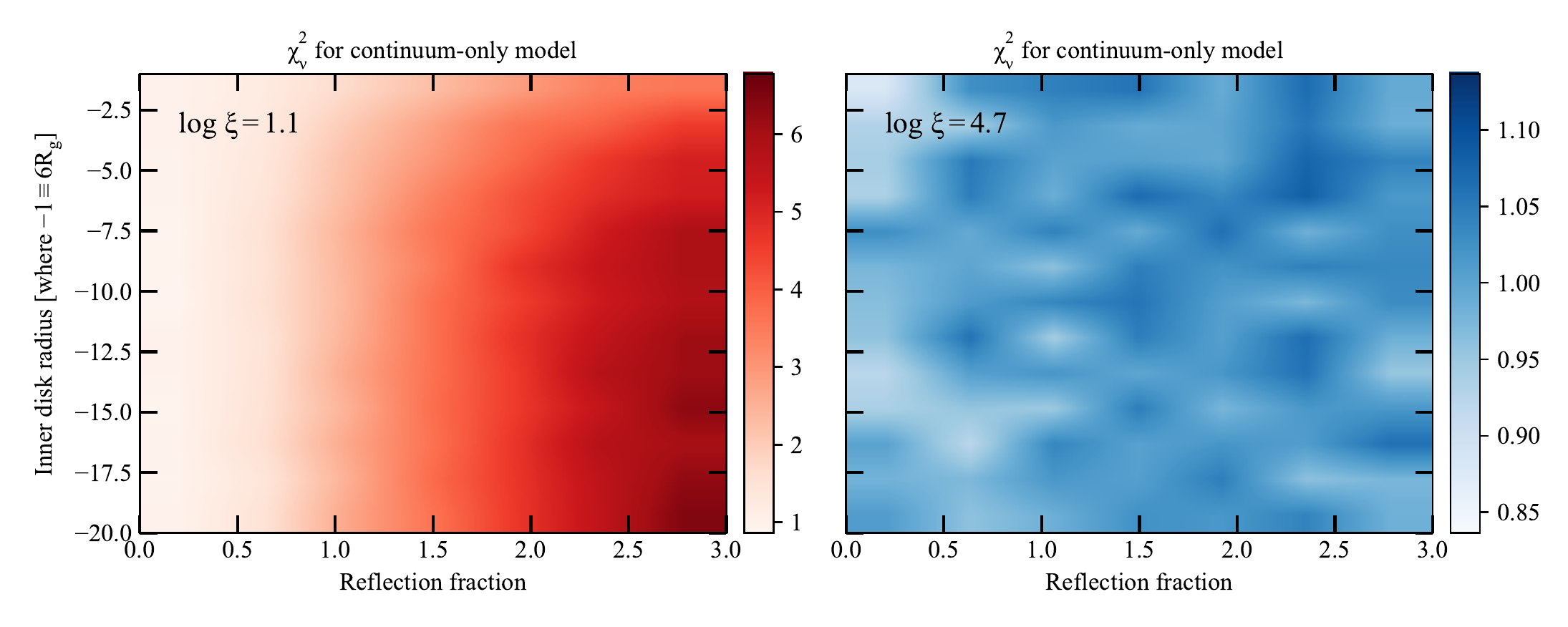}
    \caption[]{The reduced $\chi^2$ of fitting a continuum-only \textsc{tbabs}$\times$\textsc{(bbodyrad+nthcomp}) model to a simulated \textit{NuSTAR} spectrum with the same flux and exposure as the 2018 observation, containing a reflection component with the shown \textsc{relxill}-parameters. The panels show the results at low (left; $\log \xi = 1.2$) and high (right; $\log \xi = 4.5$) disk ionisation parameter. Note that, as per definition in the \textsc{relxill} model, the inner disk radius is given in units of the Innermost Stable Circular Orbit, where $-1$ equals $R_{\rm ISCO}$. The ranges in reduced $\chi^2$ differ greatly between the two ionisation states. See the main text for full details on these simulations.}
 \label{fig:sim_2panels}
\end{center}
\end{figure*}

\begin{figure}
 \begin{center}
	\includegraphics[width=\columnwidth]{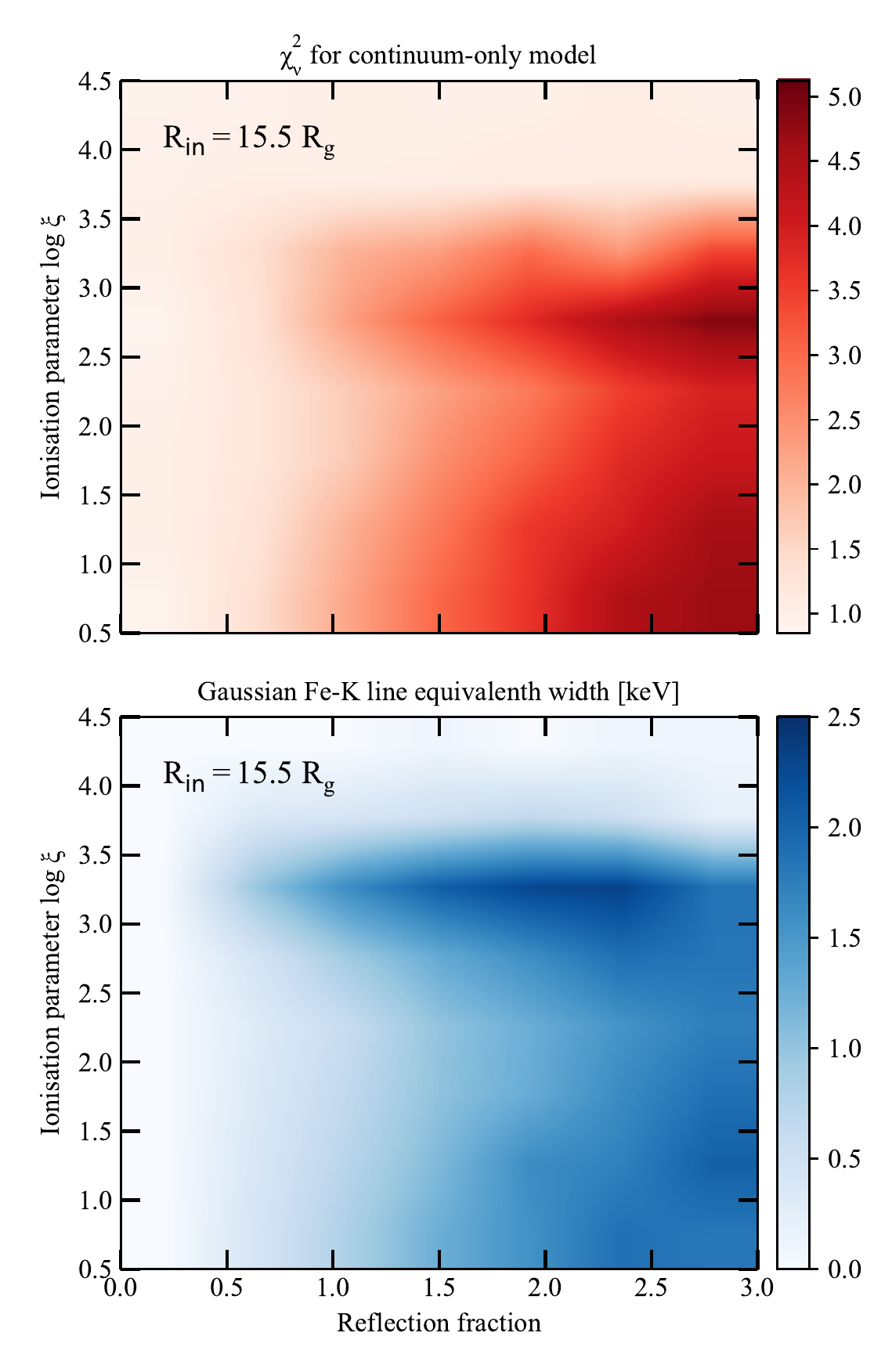}
    \caption[]{\textit{Top panels:} same as Figure \ref{fig:sim_2panels}, now showing the reduced $\chi^2$as a function of reflection fraction and disk ionisation parameter at a given inner disk radius. Reflection features become undetectable (i.e. $\chi^2_{\nu} \approx 1.0$) at low reflection fraction and very high disk ionisation. \textit{Bottom:} the equivalent width of a Gaussian iron line added to the simple continuum model, fitted to the simulated spectra. The axes are the same as the top two panels. A low equivalent width corresponds to undetectable reflection features, confirming the trends visible in the reduced $\chi^2$ diagnostic.}
 \label{fig:sim_4panels}
\end{center}
\end{figure}

%\begin{figure*}
% \begin{center}
%	\includegraphics[width=\textwidth]{1dplot_2014.pdf}
%    \caption[]{The reduced $\chi^2$ of the continuum model fitted to the synthetic \textit{NuSTAR} spectra, as a function of disk ionisation parameter, shown for low (left) and high (right) ionisation. Lines of different colors and thickness indicate different reflection fractions. As expected, for zero reflection fraction in the synthetic data, the continuum model always provide a good description. However, for non-zero reflection fractions, the reduced $\chi^2$ only approaches $1$ for the highest ionisation states ($\log \xi \gtrsim 4.5$).}
% \label{fig:sim_1d_2014}
%\end{center}
%\end{figure*}

\begin{figure*}
 \begin{center}
	\includegraphics[width=\textwidth]{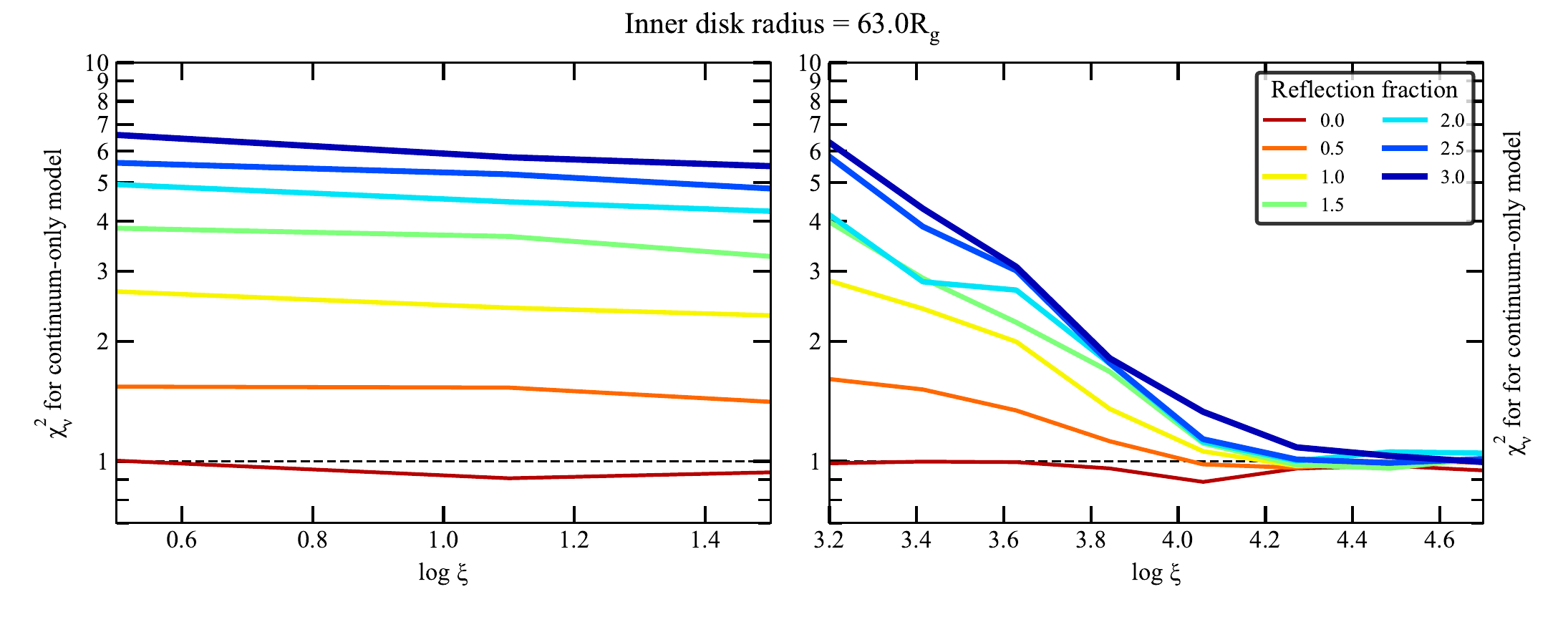}
    \caption[]{The reduced $\chi^2$ of the continuum model fitted to the synthetic \textit{NuSTAR} spectra, as a function of disk ionisation parameter, shown for low (left) and high (right) ionisation. Lines of different colors and thickness indicate different reflection fractions. As expected, for zero reflection fraction in the synthetic data, the continuum model always provide a good description. However, for non-zero reflection fractions, the reduced $\chi^2$ only approaches $1$ for the highest ionisation states ($\log \xi \gtrsim 4.1$).}
 \label{fig:sim_1d_2018}
\end{center}
\end{figure*}

Without a detected reflection signal during the 2018 \nustar\ observation, we cannot directly fit reflection parameters to determine how the accretion flow properties have changed. However, we can make use of the fact that reflection features are not detected in another way: we can instead simulate synthetic spectra with the same observational setup and flux as the 2018 observation, using different combinations of reflection parameters. For each set of reflection parameters, we can then assess whether a reflection signal would have been detected and hence constrain the changes in the accretion flow during the outburst decay, despite the absence of fittable reflection features.

Considering the relevant parameters of the \textsc{relxill} model, only three parameters can be expected to change during an outburst decay: the inner disk radius, the disk ionisation, and the reflection fraction. The remaining parameters are either continuum parameters with measured changes (power law index and normalisation), flux-independent (spin, inclination, and iron abundance), or frozen to their default values in our modelling (emissivity profile, cutoff energy). Therefore, we set up a grid in inner disk radius, ionisation, and reflection fraction, and for each combination simulate a \nustar\ spectra from the \textsc{tbabs*(bbodyrad+relxill)} model with the continuum parameters from the 2018 \nustar\ spectral shape. For each simulated spectrum at each gridpoint, we then fitted three models: the full (underlying) reflection model, a continuum-only model \textsc{tbabs*(bbodyrad+nthcomp)}, and a continuum model with an additional Gaussian Fe-K$\alpha$ line. The energy of the latter Gaussian was constrained to lie between $6.4$ and $6.96$ keV, and its equivalent width provides a diagnostic of the strength of the Fe-K$\alpha$ line in the simulated data. A second diagnostic of the strength of reflection is given by the reduced $\chi^2$ of the continuum-only model, which should approach $\sim 1.0$ as the reflection becomes undetectable. 

We performed our simulations for the following parameter grid: inner disk radius between $R_{\rm in} = 6$ and $R_{\rm in} = 120$ \rg, ionisation parameter between $\log \xi = 0.5$ and $\log \xi = 4.7$, and reflection fraction between $F_{\rm ref} = 0.0$ and $F_{\rm ref} = 3.0$. We first ran a low-resolution simulation to determine regions of interest in the probed parameter space, using 11 points in inner disk radius, 6 in ionisation parameter, and 5 in reflection fraction. Then, after concluding that the inner disk radius had little effect, while the reflection fraction and both high and low ionisation parameters did (as discussed below), we reran the simulations twice: both times with a lower resolution in inner disk radius (6 points) and a higher resolution in reflection fraction (9 points). These two added simulations used a higher resolution in ionisation parameters (6 points) to zoom in on high ($\log \xi = 3.2-4.7$) and low ($\log \xi = 0.5-1.5$) values, respectively. 

We show representative results of this grid search in Figures \ref{fig:sim_2panels} and \ref{fig:sim_4panels}: each panel shows one of the two aforementioned diagnostics of reflection strength -- reduced $\chi^2$ of the continuum-only model and Gaussian iron line equivalent width -- at a given value of one parameter, as a function of the two other parameters. In Figure \ref{fig:sim_2panels}, we show the $\chi^2_{\nu}$ of the continuum-only model at low ($\log \xi = 1.2$; left) and high ($\log \xi = 4.5$; right) ionisation. From this figure, we can draw two conclusions: first, the lack of vertical changes reveals that the value of the inner disk radius does not greatly affect the detectability of the reflection signatures. Secondly, the reflection signatures are always weak at high values of the ionisation parameter, while for lower ionisation they are weak only for low reflection fractions. 

The same results can be inferred from Figure \ref{fig:sim_4panels}. Here, the panels show the two reflection strength diagnostics for approximately the same disk inner radius. The continuum-only model fits best, again implying weak to no reflection, at low reflection fraction or high ionisation. The Gaussian Fe-K$\alpha$ line's equivalent width (bottom panel) confirms this trend, as it decreases towards those same regions of parameter space. Finally, we stress that the $\chi^2_{\nu}$ values of the continuum model shown in Figures 4 and 5 nowhere reach below the $\chi^2_{\nu}$ found when fitting the real 2018 observation with the continuum model (i.e. $\chi^2_{\nu} = 0.89$).

Finally, we show a one-dimensional representation of our simulation output in Figure \ref{fig:sim_1d_2018}. Here, we show the reduced $\chi^2$ for the simulated data fitted with the continuum-only model, as a function of disk ionisation. The lines of different colour and width represent different reflection fractions. Given the lack of any systematic effect as the inner disk radius varies (see above), we pick a representative inner radius of $63$ \rg\ and plot the results for that value. Figure \ref{fig:sim_1d_2018} confirms the inferences made above: the reflection signatures always become undetectable towards very high ionisation ( $\log \xi \geq 4.1$--$4.3$, depending on the reflection fraction). In addition, the non-detection of reflection features can be explained by a low reflection fraction ($F_{\rm ref} \approx 0$) independently of ionisation.

\section{Discussion}\label{sec:discuss}

We have presented a systematic analysis of the X-ray reflection spectrum during the outburst decay of the NS LMXB 4U 1608$-$52. Comparing \textit{Swift}, \textit{NICER}, and \textit{NuSTAR} spectra obtained over two similar outbursts, we find the following results: (i) Reflection features, such as a broad Fe-K$\alpha$ line and Compton hump, are systematically present at high mass accretion rates without changing shape, but disappear completely within days towards the end of the outburst. (ii) This non-detection of reflection features at low flux is not due to a lower signal-to-noise, but can instead be explained by either a high ionisation ($\log \xi \geq 4.1$) or low reflection fraction ($F_{\rm ref} \approx 0$) (iii) the lack of reflection features does not result from changes in inner disk radius \textit{alone}, although we stress that we cannot directly constrain the value of this radius. Here, we will discuss these results in a physical context and compare it to other LMXB systems. 

From our simulations, we conclude that the disappearance of measurable reflection features can be explained by either a very low reflection fraction, or a high disk ionisation, approaching the grid edge of the \textsc{relxill} model. The reflection fraction in the \textsc{relxill} model is defined as the ratio of the reflected and direct flux in the $20$--$40$ keV range \citep{dauser2016}. While for some flavours of reflection models, reflection fractions have been calculated self-consistently assuming given illuminating geometries, in the \textsc{relxill} model used in this work no specific geometry is assumed \citep{dauser2016}. Therefore, while it can explain the 2018 \textit{NuSTAR} spectrum, a very low reflection fraction does not directly map onto a physical picture. 

Therefore, in the remainder of the discussion, we focus more on the second explanation of a high disk ionisation, as this provides more direct physical insight into the evolution of the system. The ionisation parameter scales with the ionising X-ray flux divided by particle density, i.e. $\xi = L_{\rm X, ionising}/(n r^2)$. Comparing the two \textit{NuSTAR} spectra of 4U 1608$-$52, we find that the spectrum is harder when no reflection is detected. In other words, a higher fraction of flux comes out at high (ionising) energies. However, the total flux also decays, which counteracts this effect. More likely, an increase in ionisation parameter could be explained by a decrease in particle number density as the source decays to lower X-ray luminosities. 

What might explain such a decrease in the number density of particles in the disk? As X-ray binaries decay towards quiescence, their accretion flows are expected to change morphology. For instance, the inner accretion flow might evaporate into a radiatively inefficient hot flow \citep[for instance an advection-dominated accretion flow; ][]{narayan1994,narayan1995}. During the same transition, the inner radius of the thin disk might move outwards, truncating the Keplerian disk \citep[e.g][]{zdziarski1999,esin2001,kalemci2004,tomsick2004}. As the inner accretion flow evaporates, the density of the gas will decrease. This transition could therefore explain the increase in ionisation parameter and resulting disappearance of the reflection signatures observed in 4U 1608$-$52. The \textit{NICER} observations at higher mass accretion rates show a remarkably stable Fe-K$\alpha$ line profile, which suggest that in this scenario, the disk initially remains stable. Subsequently, the disk changes morphology within days, as it evaporates into a radiatively inefficient hot flow below $\sim 10^{36}$ erg $s^{-1}$ (i.e. below $\sim 10^{-2}$ \ledd). Due to gaps in the X-ray monitoring and required grouping of the \textit{NICER} observations, the exact transition X-ray luminosity cannot be measured precisely. 

The possible transition to a radiatively inefficient hot flow before the \textit{NuSTAR} observation in the 2018 outburst can be compared to the behaviour of other neutron star LMXBs. For instance, the continuum blackbody and comptonization parameters during the 2018 \textit{NuSTAR} observation (e.g., Table \ref{tab:2018spectrum}) are very similar to those observed for Aql X-1 with \suzaku\ at a similar luminosity level \citep[][]{sakurai2014}. At this luminosity, no reflection features are detected in Aql X-1 either, which might similarly arise from the evaporation into a hot flow. It should be noted, however, that reflection features in Aql X-1 are also weak at higher luminosity and that the disk is likely always truncated away from the neutron star \citep{king2016,ludlam2017_aqlx1}. Therefore, the inner disk geometry might not be exactly comparable and the reflection spectrum could simply be undetectable due to the low flux. This comparison with Aql X-1 highlights the exceptional combination of strong reflection and low distance presented by 4U 1608$-$52, which allows us to conclude that the lack of reflection does not result only from limited signal-to-noise. 

An additional comparison can be made with very-faint X-ray binaries (VFXBs), which never reach high accretion luminosities, as they either persistently accrete between $\sim 10^{34}$ to $10^{36}~\lum$ or exhibit outbursts that peak below $10^{36}~\lum$. Of the few VFXBs with detailed spectral observations, most do not show reflection features \citep[e.g.,][]{armas2011,armas2013,armas2013_2,lotti2016,sanna2018_2,sanna2018}. Moreover, the transient black hole LMXB Swift J1357.2-0933, which does not become brighter than $\sim 10^{35}~\lum$ during its accretion outbursts, does not show an Fe-K$\alpha$ line either \citep[][]{armas2014,beri2019}. Being in the same X-ray luminosity range as 4U 1608$-$52 during the 2018 \textit{NuSTAR} observation, the lack of reflection in these VFXBs might also be explained by a radiatively inefficient accretion flow with a high ionisation parameter. 

Interestingly, weak reflection features were detected with \nustar, \xmm\ and \chan\ in the (persistently accreting) neutron star VFXB IGR J17062$-$6143. Reflection modelling of those data implied that the disk could be truncated at $\sim 100$ \rg~\citep[][]{degenaar2017_igrj1706,vandeneijnden2018_igr}. In this source, and generally in neutron star X-ray binaries showing truncated disks, it is difficult to distinguish between two possible explanations for the disk truncation: a relatively strong magnetic field, or the evaporation of the inner thin disk. However, our results for 4U 1608$-$52 might provide some insight: in IGR J17062$-$6143, reflection is detected despite a strongly truncated accretion flow. Based on our results on 4U 1608$-$52, we can envision three scenarios: firstly, that the disk in IGR J17062$-$6143 is also evaporated but has not reached high enough ionisation levels to yield reflection features undetectable; secondly that the disk truncation is instead caused by a different mechanism, such as the neutron star magnetosphere truncated an otherwise relatively thin disk. The presence of X-ray pulsations, making IGR J17062$-$6143 an accreting millisecond X-ray pulsar \citep{strohmayer2017,strohmayer2018}, plausibly fits within this picture, as matter is channeled to the magnetic poles; or thirdly, that the disappearance of reflection features in 4U 1608$-$52 is caused by a low reflection fraction instead, which doesn't occur in IGR J1706-2$-$6143.
%if a neutron star system with an evaporated inner thin disk indeed will not show detectable reflection features, the presence of reflection in IGR J17062$-$6143, albeit weak, could suggest that the measured disk truncation is instead caused by a different mechanism, such as the neutron star magnetosphere. The presence of X-ray pulsations, making IGR J17062$-$6143 an accreting millisecond X-ray pulsar \citep{strohmayer2017,strohmayer2018}, plausibly fits within this picture, as matter is channeled to the magnetic poles. However, we again stress that these inferences only hold if the disappearance of reflection features in 4U 1608$-$52 results from an increased ionisation, instead of a decreased reflection fraction.

Here, we briefly consider the effect of the continuum shape on the detectability of reflection features. Analysing four \textit{Suzaku} observations of 4U 1608$-$52 throughout its 2010 outburst, \citet{armas2017} find no evidence for reflection at any moment, placing limits on the equivalent width of the Fe-K$\alpha$ line of $<14$~eV. These spectra were taken in the range of $\sim (1-20) \times 10^{36}~\lum$, a factor two to forty brighter than the 2018 \textit{NuSTAR} spectrum. \citet{armas2017} note that this lack of reflection could result from either an intrinsic lack of reflection features, or burying by the continuum. We find that at lower luminosity it can be a combination of both: while reflection should have been seen in the 2018 \textit{NuSTAR} observation if it were present (Section \ref{sec:results_isitburied}), it becomes undetectable at lower ionisation parameter for the 2018 continuum shape. However, the comparison between \textit{NuSTAR} and \textit{Suzaku} is more accurate using the 2014 observation, which overlaps in flux with the \textit{Suzaku} range \citep{degenaar2015_4u1608}. Given the strong reflection in the \textit{NuSTAR} spectrum, and in the \textit{NICER} monitoring around the 2018 outburst peak, a low signal-to-noise seems the most likely explanation of the lack of reflection detected by \citet{armas2017}.

If reflection had been detected in 4U 1608$-$52 during the 2018 \textit{NuSTAR} observation, we could have measured the disk inner radius and compared it with the earlier \textit{NuSTAR} measurement. However, in the absence of reflection, we do not obtain any constraints on the inner disk radius. While it could very well change in tandem with the changing ionisation state, our simulations show that varying the inner disk radius alone cannot explain the disappearance of reflection features. Therefore, we cannot place any constraints on the evolution or value of \rg in the 2018 observation. 

Finally, in the few black hole LMXBs observed at similarly low X-ray luminosities, the inner disk appears to be truncated: in GX 339$-$4, the inner disk radius appears to move from a few gravitational radii at high accretion rate to $\gtrsim35$~\rg\ at $\sim 10^{-3}$~\ledd\  \citep[][]{tomsick2009, garcia2019}. A \nustar\ observation of GRS 1739$-$278 places its inner disk at $\sim15-35$~\rg\ when accreting at $\sim 10^{-2}$~\ledd\ \citep[][]{fuerst2016}. Interestingly, in these sources, the Fe-K$\alpha$ line is clearly detected. In the scenario proposed above, where we assume ionisation changes caused by a transition to a  low-density, radiatively inefficient accretion flow drive the iron line disappearance, instead of low reflection fraction, this FeK$\alpha$-line is not expected. Therefore there might be a difference in the evaporation of the disk as the thin disk recedes in these two black hole systems, when compared to 4U 1608$-$52. 

\section*{Acknowledgements}
The authors thank the anonymous referee for comments that improved the quality of this manuscript. JvdE, ND and ASP are supported by an NWO Vidi grant awarded to ND. RML acknowledges the support of NASA through Hubble Fellowship Program grant HST-HF2-51440.001. PB was supported by an NPP fellowship at NASA Goddard Space Flight Center. This research has made use of the MAXI data provided by RIKEN, JAXA and the MAXI team, together with data and software provided by the High Energy Astrophysics Science Archive Research Center (HEASARC) and NASA's Astrophysics Data System Bibliographic Services. This work made use of data from the NuSTAR mission, a project led by the California Institute of Technology, managed by the Jet Propulsion Laboratory, and funded by the National Aeronautics and Space Administration. This research has made use of the NuSTAR Data Analysis Software (NuSTARDAS) jointly developed by the ASI Science Data Center (ASDC, Italy) and the California Institute of Technology (USA). This work was supported in part by NASA through the \textit{NICER} mission and the Astrophysics Explorers Program.

%\footnotesize{
%\bibliographystyle{mn2e}
%\bibliography{thesis}

%}

\appendix

\section{Results from XRT spectral fitting}

Table \ref{tab:swiftspec} lists the details for our Swift XRT spectral fits to measure the flux evolution throughout the 2018 outburst of 4U 1608$-$52.

\label{app:XRT}
\begin{table*}
\caption{Results from the \swift/XRT spectral continuum fits to measure the X-ray flux evolution of 4U 1608$-$52. \label{tab:swiftspec}}
\begin{threeparttable}
\begin{tabular*}{1.0\textwidth}{@{\extracolsep{\fill}}lccc}
\hline
\multirow{2}{*}{MJD} & \multirow{2}{*}{ObsId} & \multirow{2}{*}{$\Gamma$}& $F_{\mathrm{X}}$ \\
& & & ($\flux$) \\
\hline
58302.4540 & 00010741001 & 1.90 $\pm$ 0.01 & $(7.81 \pm 0.03) \times 10^{-9}$ \\ %  & XX (XX)  \\
58304.6494 & 00010741002 & 2.17 $\pm$ 0.01 & $(3.55 \pm 0.02) \times 10^{-9}$ \\ %&  XX (XX) \\
58306.4631 & 00010741003 & 2.27 $\pm$ 0.01 & $(2.76 \pm 0.02) \times 10^{-9}$  \\ %&  XX (XX)  \\
58308.2397 & 00010741004 & 2.27 $\pm$ 0.02 & $(8.99 \pm 0.10) \times 10^{-10}$   \\ %&  XX (XX)  \\
58310.5458 & 00010741005 & 2.28 $\pm$ 0.05 & $(7.07 \pm 0.15) \times 10^{-10}$   \\ %&  XX (XX)  \\
58312.2109 & 00010741006 & 2.21 $\pm$ 0.10 & $(2.90 \pm 0.15) \times 10^{-10}$  \\ %&  XX (XX)  \\ 
58314.0427 & 00010741007 & 2.06 $\pm$ 0.09 & $(1.70 \pm 0.08) \times 10^{-10}$   \\ %& XX (XX)  \\
58316.4016 & 00010741008 & 1.84 $\pm$ 0.10 & $(6.74 \pm 0.32) \times 10^{-10}$   \\ %&  XX (XX)  \\
58318.3281 & 00010741009 & 1.95$\pm$ 0.09 & $(4.96 \pm 0.23) \times 10^{-10}$   \\ %& XX (XX)  \\
58320.3856 & 00010741010 & 1.87 $\pm$ 0.08 & $(6.05 \pm 0.25) \times 10^{-10}$   \\ %&  XX (XX)  \\
58322.5791 & 00010741011 & 1.85 $\pm$ 0.09 & $(4.98 \pm 0.22) \times 10^{-10}$   \\ %& XX (XX)  \\
58326.7577 & 00010741012 & 2.04 $\pm$ 0.02 & $(8.80 \pm 0.10) \times 10^{-10}$   \\ %&  XX (XX)  \\
58330.6144 & 00010741013 & 1.73 $\pm$ 0.02 & $(9.51 \pm 0.10) \times 10^{-10}$   \\ %&  XX (XX)  \\
58334.2006 & 00010741014 & 1.81 $\pm$ 0.03 & $(9.34 \pm 0.12) \times 10^{-10}$   \\ %& XX (XX)  \\
58338.0529 & 00010741015 & 1.89 $\pm$ 0.10 & $(6.93 \pm 0.34) \times 10^{-10}$   \\ %&  XX (XX)  \\
58348.6161 & 00010741016 & 1.76 $\pm$ 0.14 & $(7.06 \pm 0.05) \times 10^{-11}$   \\ %&  XX (XX)  \\
58353.0956 & 00010741017 & 2.50 $\pm$ 0.61 & $(3.88 \pm 1.34) \times 10^{-12}$   \\ %&  XX (XX)  \\
58360.5705 & 00010741018 & 3.66 $\pm$ 1.05 & $(3.05 \pm 3.73) \times 10^{-12}$   \\ %&  XX (XX)  \\
58363.0151 & 00010741019 & 4.01 $\pm$ 1.14 & $(3.37 \pm 3.84) \times 10^{-12}$   \\ %&  XX (XX)  \\
58367.4035 & 00010741020 & 1.87 $\pm$ 1.03 & $(1.79 \pm 1.01) \times 10^{-12}$   \\ %&  XX (XX)  \\
58381.9211 & 00000000000 & 4.32 $\pm$ 1.07 & $(3.10 \pm 2.91) \times 10^{-12}$  \\ %&  XX (XX)  \\
58415.8456 & 00010741026 & 3.59 $\pm$ 1.42 & $(2.62 \pm 3.62) \times 10^{-12}$   \\ %&  XX (XX)  \\
\hline
\end{tabular*}
\begin{tablenotes}
\item[]Note -- These spectral fits were performed with $N_{\mathrm{H}}=1.58\times 10^{22}~\nh$ fixed. $F_{\mathrm{X}}$ represents the unabsorbed 0.5--10 keV model flux. The model used was \textsc{tbabs*pegpwrlw}.
\end{tablenotes}
\end{threeparttable}
\end{table*}

\end{document}